\documentclass[conference,a4paper]{IEEEtran}

\makeatletter
\def\ps@headings{%
\def\@oddhead{\mbox{}\scriptsize\rightmark \hfil \thepage}%
\def\@evenhead{\scriptsize\thepage \hfil \leftmark\mbox{}}%
\def\@oddfoot{}%
\def\@evenfoot{}}
\makeatother
\usepackage{graphicx}
\usepackage{amssymb,amsmath,latexsym,amsfonts}
\usepackage{epsfig}
\usepackage{comment}

\newcommand{\be}{\begin{equation}}
\newcommand{\bel}[1]{\begin{equation}\label{#1}}
\newcommand{\ee}{\end{equation}}
\newcommand{\ba}{\begin{align}}
\newcommand {\N} {{\rm I\kern-2.5pt N}}
\newcommand{\ea}{\end{align}}

\newtheorem{proposition}{Proposition}

\newtheorem{remark}{Remark}

\begin{document}

\title{Predicting the Impact of Measures Against P2P Networks
on the Transient Behaviors}


\begin{comment}
\author{\authorblockN{Eitan Altman, Philippe Nain, Adam Shwartz, Yuedong Xu}
\authorblockA{INRIA Sophia Antipolis\\
BP 93, 06902 Sophia Antipolis, France\\
Email: \{eitan.altman, yuedong.xu, phinippe.nain\}@inria.fr} and adam@ee.technion.ac.il
%\and
%\authorblockN{Philippe Nain}
%\authorblockA{INRIA Sophia Antipolis\\
%Email: eitan.altman@sophia.inria.fr}
%\and
%\authorblockN{Eitan Altman}
%\authorblockA{INRIA Sophia Antipolis\\
%Email: philippe.nain@sophia.inria.fr}}
}
\end{comment}

\author{\authorblockN{Eitan Altman, Philippe Nain}
\authorblockA{INRIA Sophia Antipolis \\
2004 Route des Lucioles\\
\{eitan.altman,philippe.nain\}@inria.fr}
\and
\authorblockN{Adam Shwartz}
\authorblockA{Technion \\
Haifa 32000, Israel\\
adam@ee.technion.ac.il}
\and
\authorblockN{Yuedong Xu}
\authorblockA{INRIA Sophia Antipolis \\
2004 Route des Lucioles \\
yuedong.xu@inria.fr}
}

\maketitle
%\vspace{-10cm}

\begin{abstract}
The paper has two objectives. The first is to study
rigorously the transient behavior of some peer-to-peer (P2P)
networks whenever information is replicated and disseminated
according to epidemic-like dynamics.
The second is to use the insight gained from the previous analysis in order to predict
how efficient are measures taken against P2P networks.
We first introduce a stochastic model which extends a classical epidemic model, and
characterize the P2P swarm behavior in presence of free riding peers.
We then study a second model in which a peer initiates a contact with another peer
chosen randomly. In both cases the network
is shown to exhibit phase transitions: a small change in the
parameters causes a large change in the behavior of the network.
We show, in particular, how phase transitions affect measures of content providers
 against P2P networks that distribute non-authorized music or books, and what is the
 efficiency of counter-measures.
\begin{comment}
In both cases,
the network is shown to have a  phase transition which can
be exploited by the measures that the content provider industry
may take against P2P networks.

We explicitly study
how system's parameters affect the efficiency of such measures.
\end{comment}
\end{abstract}


% for peerreview papers, inserts a page break and creates the second title.
% Will be ignored for other modes.
\IEEEpeerreviewmaketitle

\section{Introduction}
\label{sec:intro}

Along with the worldwide penetration of the Internet,
a huge demand has appeared to copyrighted music and movies that
have been accessible for free over the Internet.
While benefiting a very large internaut community as well
as potentially providing higher benefits for Internet access providers,
it seems unclear whether the creators and the copyright owners have
gained anything from this unregulated access.
Two opposing approaches appeared, both proposing to protect
the copyright owners. The first consists of fighting against
non-authorized access whereas
the second one, aims at finding  cooperative solutions
that would benefit both the Internauts as well as all other
economic actors. An example of a cooperative solution is
a flat taxation that would allow Internauts to pursue downloading
freely music and films, and that would distribute the tax money
between the copyright owners. This cooperative approach has several
difficulties in its implementation; a major one is
how to distribute the tax income fairly.
A major drawback of the confrontation policy is the huge monitoring
effort that it requires and that seems not to provide credible
evidence for unauthorized downloads \cite{printer}.
In order to assess the efficiency of non-cooperative measures against
unauthorized downloads, the authors of
\cite{sulan1} have analyzed
the impact of the effort, of the authorities or of content provider
companies, invested
in (i) reducing file uploading in P2P networks and in (ii)
reducing the demand for files, on the availability of files
and, more generally, on the operation of the P2P
networks. The stationary analysis there is based on a M/G/$\infty$
queuing model.

In this paper we are interested in predicting the impact of measures as described in the
previous paragraph, on the {\em transient} behavior of torrents.
By how much should the request or departure rate in a P2P network be reduced
in order to have a significant change in file availability? To achieve that, we
consider abstract models of a torrent in simplified P2P networks,
where a large number of peers are interested in a file which is initially
available at a small fraction of the population.

Our models are  formulated as epidemic type processes of file dissemination.
We consider both {\em cooperative} peers, which are those that
make a file available to other peers as soon as they obtain the
file, and  {\em free riders}, who leave the system immediately
after obtaining the file.
To understand the impact of measures against the cooperative sharing
behavior, we parameterize the degree of free-riding in the system
as well as the degree of cooperation.

The P2P dynamics is modeled by a Markov chain
(Section \ref{sec:model}) which is approximated in two specific
regimes: the first (Section \ref{sec:branch}) is
the early stage when a large fraction of
the population does not yet have the file.
The system is then well approximated by a branching process.
In the case that there is
a positive probability of not getting extinct in the first regime, the system
is shown in Section \ref{sec:fluid} to move with some non-zero probability
to a second regime in which, for the case of a sufficiently
large population size, its dynamics is close to the solution of a
differential equation.
A similar fluid limit is studied in \ref{sec:modelSi} for the case
of limitation on uplink or downlink speed.
We briefly state our contributions:

{\bf 1. Modeling and approximating the transient behavior}
Our first important contribution is to show in what sense each of the
above two models approximates the original Markov chain, and how to
use both in order to get the whole transient
behavior of the P2P network.
This is in contrast with all other models of P2P networks
that we know  of, which either use only a branching process approach
\cite{Infocom04:Yang}
or which use only an epidemic mean-field approximation
\cite{Sigcomm04:Qiu}. The latter approach
(of using only the mean-field limit) is shown to provide a tight approximation
when the initial number of peers with the file scales linearly
with the total size $N$ of the population of  peers. With a fixed initial number
of nodes that does not scale with $N$, there is a positive
probability of early extinction (see Section \ref{sec:impact} for detail) for any set of system parameters,
and this probability cannot be predicted by the mean-field limit alone.

{\bf 2. Analysis and identifying phase transitions}
We first study a P2P model that corresponds to the
epidemic-like file dissemination (Sections \ref{sec:model}-\ref{sec:fluid})
We then study a second model (Section \ref{sec:modelSi}) in which, at random times, each peer contacts another peer randomly chosen
within the set of  existing peers.
In both cases, we show the existence of phase transitions: a small change
in the parameters causes a large change in the network behavior.

A phase transition occurs both in the branching model for the
extinction time  and in the epidemic model for the file availability.
In the branching process, the existence of
two phases was not known  to Galton and Watson (considered as the
founders of branching processes) and was only discovered
and proved later in \cite{stef1933}. In the epidemiology community,
the phase transition was already
known  in \cite{epid} for a model equivalent to our first  model
without the free riders.
For the the second model \cite{Si}, we show the existence
of two phase transitions, one  for the file availability and the other one for the maximum torrent
size.

{\bf 3. Application.}
In Section \ref{sec:control}, we present  a counteraction against
unauthorized file sharing in the presence of illegal publishers.
We evaluate the impact of measures against Internet piracy on the performance
of P2P systems in Section \ref{sec:impact} (see Figure \ref{fig:control1}).

The accuracy of the various approximations is investigated in Section \ref{sec:impact}, related studies
are discussed in Section \ref{sec:related}, and concluding remarks are given in  Section \ref{sec:conclude}.

\section{Model}
\label{sec:model}
\subsection{Assumptions}
\label{ssec:ass}
Assume there is a population $N$ of peers interested in a single file.
Let $Y(t)$ be the number of peers that possess the file at time $t$.
A peer acquires the file when it encounters
another peer that has the file. We will consider two types of peers:  {\em cooperative} and {\em non-cooperative}
peers. Once a cooperative peer has acquired the file, it stays in the network for a random time distributed
according to an exponential rv with parameter $1/\mu\geq 0$ and then leaves the network. During the lingering time of
a cooperative peer with the file, it participates in the file dissemination.
A non-cooperative peer, also called a {\em free-rider}, leaves the network at once when it receives the file.
Note that ``free riders'' in our context is an abstract description of noncooperative behaviors, which is
different from that in the current BitTorrent system.

Let $X_c(t)$ and $X_f(t)$ denote the number of
cooperative peers without the file and the number of free-riders (necessarily without the file) at time $t$, respectively.
Define the process ${\bf Y}:=\{(Y(t),X_c(t),X_f(t)),t\geq 0\}$.
Let $(Y(0), X_c(0),X_f(0))$ denote the initial state of ${\bf Y}$ that has $Y(0)+X_c(0)+X_f(0)=N$.
%We denote by $r$ the ratio of cooperative peers that has $r = (Y(0)+X_c(0))/N$.
Let the ratio of various types of peers be $(y_0, x_{c,0}, x_{f,0}) := (\frac{Y(0)}{N},\frac{X_c(0)}{N},\frac{X_f(0)}{N})$.
For simplicity, we introduce new
variables $N_c = X_c(0)$ and $N_f=X_f(0)$.

%We denote by $x_{c,0}:=N_c$ the number of cooperative peers without
%the file at time $t=0$. We further assume that $y_0 \in \{1,\ldots, N-1\}$
%($y_0=0$ implies that $Y(t)=0$ for all $t$ and $y_0=N$ implies that $X_c(t)=X_f(t)=0$ for all $t$, two
%cases which have no interest). Let $r$ denote the ratio of cooperative peers that has $r = (y_0+x_{c,0})/N$.

We consider an abstract P2P network in which the file acquisition is via random contact between pair-wise peers.
When two such peers meet, the cooperative peer transmits the file to the other peer.
It is assumed that it takes an exponential time with rate $\lambda>0 $ for a peer without a file to encounter a
cooperative peer with the file. The transmission of the file is always supposed to be successful.
This model describes a general P2P swarm without a tracker, and even the spreading of a file in current Internet.
It is inspired by the contact process in \cite{Si} and \cite{Sigmetrics05:Massoulie}. One of the main difference
lies in that a peer contacts all other connected peers in the system, instead of only one random peer periodically.
%The random contact also characterizes the important property of \emph{flash crowds}.
We assume that the file transmission time is negligible
compared to the time it takes for two peers to meet and therefore this time is taken to be zero.

\begin{comment}
The random contact also characterizes the important property of \emph{flash crowds}.
When two such peers meet, the cooperative peer transmits the file to the other peer.
It is assumed that it takes an exponential time with rate $\lambda>0 $ for a peer without a file to encounter a
cooperative peer with the file. %When two such peers meet, the cooperative peer transmits the file to the other peer.
The transmission of the file is always supposed to be successful. We assume that the file transmission time is negligible
compared  to the time it takes for two peers to meet and therefore this time is taken to be zero.
\end{comment}

All the random variables (rvs) introduced so far are assumed to be mutually independent. As a consequence, if $Y(t)=k$
then any peer without the file will meet a cooperative peer with the file
after a time that is distributed according to the minimum of $k$ independent and exponential rvs with rate
$\lambda$, that is after a time distributed according to an exponential rv with rate $\lambda k$.

Measures of the authorities or of content provider companies
against file sharing systems may have an impact on %A typo
the decrease in the population $N$ interested in the file and
an increase in the fraction of free riders among the population
interested in the file. It can however have an impact also on
the behavior of cooperative peers that would leave the %A syntax
system sooner (i.e.\ $\mu$ is expected to increase). %A syntax
Our model combines an epidemic type propagation
of the file together with a description of the free riding
behavior. Define  $\rho:=\lambda N_c/\mu$.

We first consider (Section \ref{sec:fullycoop}) the case where all peers are fully cooperative in the sense that $\mu=0$
and $X_f(0)=0$ (no free riders).  $\mu=0$ implies that cooperative peers do not leave the network
after receiving the file. We then move to the general case where $\mu>0$ and $X_f(0)\geq 0$ (Section \ref{sub-general-network}).

\subsection{Fully cooperative network}
\label{sec:fullycoop}
When all peers are fully cooperative (i.e. $\mu=0$ and $X_f(0)=0$) the population of peers remains
constant and equal to $N$, that is, $Y(t)+X_c(t)=N$ at any time $t$. The network dynamics can be
represented by the process $\{Y(t),t\geq 0\}$.

This is a finite-state continuous-time Markov
process with non-zero transitions given by
{\small
\begin{equation}
\label{e:closed1}
Y (t) \to Y (t)+1 \quad \mbox{ with rate} \quad \lambda Y(t) (N-Y(t)).
\end{equation}
}
In other words the process $\{Y(t),t\geq 0\}$  is a pure birth Markov process on the state-space
$\{y_0,\ldots,N\}$, where state $N$ is an absorbing state which is reached when all
peers have the file.

Define $m(t):=E[Y(t)]$, the expected number of peers with the file at time $t$. Standard algebra shows that
\begin{equation}
\label{epidemics-stochastic}
\frac{d m(t)}{dt}=\lambda E[Y(t)(N-Y(t))],\quad t>0.
\end{equation}
Unfortunately, the right-hand side of (\ref{epidemics-stochastic}) does not express as a function of $m(t)$,
thereby ruling out the possibility of finding $m(t)$ in closed-form as the solution of an ODE.

Assume that $\lambda$ is written as $\lambda = \beta/N$ and that $\lim_{N\to\infty} N^{-1}Y(0)= y_0\in (0,1]$.
Then, for large $N$, $m(t)$ is well-approximated by $N y(t)$ where $y(t)$ is obtained
as the unique solution of the ODE \cite[Thm 3.1]{Kurtz}
\begin{equation}
\label{epidemics-deterministic}
\frac{dy(t)}{dt}= f(y(t)), \quad t>0,
\end{equation}
where $f(u):=\beta u(1-u)$ and $y(0)=y_0 \in (0,1]$ (conditions (3.2)-(3-4) in \cite[Thm 3.1]{Kurtz} are clearly satisfied). It is found that
\begin{equation}
\label{ode-y}
y(t)= \frac{y_0}{(y_0+(1-y_0)\exp(-\beta t))}, \quad t\geq 0.
\end{equation}

This is a well-known instance (see e.g. \cite{Umass-epidemics}) of what is known as mean-field approximation,
a theory that focuses on the solution of ODEs obtained as limits of jump Markov processes \cite{Kurtz}.
The ODE (\ref{epidemics-deterministic}) has been extensively used in epidemiology studies, where $y(t)$ represents
the fraction of infected patients at time $t$ when the population is of size $N$.

Proposition  \ref{theorem:mfa_bound} below, whose proof can be found in \cite{TR10},
states  that  the mean-field approximation is an upper bound for $E[Y(t)]$.
\begin{proposition}%\footnote{All the proofs can be found in the technical report \cite{TR10}.}
\label{theorem:mfa_bound}
 $E[Y(t)]\leq Ny_0 /(y_0+(1-y_0)e^{-\beta t})$  $\forall t\geq 0$.
\end{proposition}

\subsection{General network}
\label{sub-general-network}
We consider the general network defined in Section \ref{ssec:ass}.

Define the vector  $X(t)= { X_c(t) \choose X_f(t) } $, where we recall that
$X_c(t)$ is the number of cooperative nodes in the system
who do not have the file at time $t$ and  $X_f(t)$ is the  number
of free-riders in the system at time $t$ (by definition, none of
these have the file at time $t$). Let $e_c=(1,0)$ and $e_f=(0,1)$.
Under the statistical assumptions made in Section \ref{ssec:ass}
it is seen that the process ${\bf Y}=\{(Y(t),X(t)),\,t\geq 0\}$ is a finite-state Markov process whose non-zero
transitions are given by
{\small
\bel{e:closed1b1}
\begin{pmatrix}
Y (t) \\ X(t)
\end{pmatrix}
\to
\begin{pmatrix}
Y (t)+1  \\ X (t)- e_c
\end{pmatrix}
\mbox{ with rate $\lambda Y (t) X_c (t) $,}
\ee
\bel{e:closed1b2}
\begin{pmatrix}
Y (t) \\ X(t)
\end{pmatrix}
\to
\begin{pmatrix}
Y (t) -1 \\ X(t)
\end{pmatrix}
\mbox{ with rate $\mu Y(t) $,}
\ee
\bel{e:closed1b3}
\begin{pmatrix}
Y (t) \\ X(t)
\end{pmatrix}
\to
\begin{pmatrix}
Y (t) \\ X(t)  - e_f
\end{pmatrix}
\mbox{ with rate $\lambda Y(t) X_f (t) $.}
\ee
}
Throughout this paper we will assume that $\lambda>0$ and $\mu>0$.

The process {\bf Y} takes its values in the set
${\cal E}:=\{(i,j,k), 0\leq i\leq y_0+N_c, 0\leq j\leq N_c, 0\leq j+k\leq N-y_0\}$.
Furthermore, all states in ${\cal E}$ of the form $(0,j,k)$ are absorbing states since there are no more
transitions when the file has disappeared.

%Note that the process ${\bf Z}:=\{(Y(t), X_c(t)),\,t \geq 0\}$ is also an absorbing  finite-state Markov process
%whose non-zero transitions are given in (\ref{e:closed1b1})-(\ref{e:closed1b2}).

An explicit characterization of the transient behavior of the absorbing Markov process {\bf Y} is a difficult
task due both to the presence of non-linear and non-homogeneous transition rates in the state variables and
to the dimension of ${\bf Y}$. %The same is true for ${\bf Z}$.
\begin{comment}
If $N$ is not too large, a numerical approach is possible. We refer the interested reader to \cite{IAN} where this
approach was used in a related context.
\end{comment}
In this paper we will instead develop two approximations of the Markov process ${\bf Y}$.
The first one, in Section \ref{sec:branch}, will consist in replacing $X_c(t)$ by $N_c=X_c(0)$ in the transition rate
(\ref{e:closed1b1}), which will introduce a birth and death Markov branching process. As expected,
this (so-called) branching approximation will loose its accuracy as the ratio
$X_c(t)/N_c$ decreases.

The second approximation, in Section \ref{sec:fluid},  will use an asymptotic argument as $N\to\infty$ based on a mean-field
approximation of  {\bf Y}. This approximation is justified if  the initial state of ${\bf Y}$
is of the order of $N$.
Both the branching and the mean-fielf approximations
approaches will allow us to approximate key characteristics of  {\bf Y} such as the
probability of disappearance of the file, the time before all files disappear, the maximum number
of cooperative peers in the network and the fraction of peers that eventually receive the file.
\begin{comment}
\begin{remark}
Our general model is inspired by the one in \cite{Si}. Let us briefly point out some differences between our model
and the one in \cite{Si}.
\begin{itemize}
\item
The model of \cite{Si} assumes that cooperative users
remain only a finite time in the system as a free-riding
phenomenon and thus calls $\mu$ a measure for free-riding.
Our model has in addition the parameter $\eta:=N_c/(N-y_0)$ that quantifies
free riding. This motivated us to define $\eta$ as a free riding
measure and $\mu$ as a measure of cooperation.
\item
The model for interaction between nodes is different. %A changed this item - was not clear
This results in a different rate of increase of the number
of nodes with the file.
\item
In \cite{Si} there are some seeders that always remain in the system.
Therefore unlike our case, in \cite{Si} all peers eventually receive
the file.
\end{itemize}
\end{remark}
\end{comment}

\section{Branching approximation}
\label{sec:branch} %A added label

Let  ${\bf Y}_{b}:=\{Y_{b}(t),\,t\geq 0\}$ be a Markov process on $\N:=\{0,1,\ldots\}$
(the subscript $b$ refers to ``branching'')
with non-zero transition rates given by
{\small
\begin{align}
Y_{b} (t) & \to Y_{b} (t) + 1 & \mbox{with rate $\lambda Y_{b}  (t) N_c $}
\label{Yu-birth}\\
Y_{b} (t) & \to Y_{b} (t) - 1 & \mbox{with rate $\mu Y_{b} (t) $}
\label{Yu-death}
\end{align}
}
where we recall that $N_c$ is the number of cooperative peers without the file at time $t=0$.

Since $X_c(t)$, the number of cooperative peers without the file at time $t$, is non-increasing
in $t$, a quick comparison  between (\ref{e:closed1b1})-(\ref{e:closed1b3}) and (\ref{Yu-birth})-(\ref{Yu-death})
indicates that the process ${\bf Y}_{b}$  should dominate the process  ${\bf Y}$. This
bounding result is formalized and proved in the proposition below.

A word on the notation: a real-valued rv $Z_1$ is stochastically smaller than another real-valued rv $Z_2$, denoted
as $Z_1\leq _{st} Z_2$, if $P(Z_1>x)\leq P(Z_2>x)$ for all $x$.

\begin{proposition}
\label{prop:bound}
If $Y(0)\leq Y_{b}(0)$ then $Y(t)\leq _{st} Y_{b}(t)$ for any $t>0$.
%(ii) As $N_c$ becomes large, the approximation becomes tight.
\end{proposition}
%\noindent \textbf{Proof.} Please refer to the technical report \cite{TR10}.\done
%\vskip0.3true cm
The Markov process ${\bf Y}_{b}$ is an absorbing continuous-time birth and death process on $\N$  with absorbing state $0$. Because its
transition rates are linear functions of the system state, this is also a continuous-time Markov {\em branching process} \cite{Ha}, namely,
a process in which at any time $t$ each member of $Y_{b}(t)$ evolves independently of each other.
The next section specializes known results of the theory of branching processes to the process ${\bf Y}_{b}$.

\subsection{Extinction probability and extinction time}

As previously observed the process ${\bf Y}_b$ is a birth and death branching process  \cite[Chapter V]{Ha}.
Each object (peer) of this process has a probability of change in the interval $(t,t+h)$ given by $b h  + o(h)$
with $u = \lambda N_c+ \mu$; with probability $p_0=\mu /u$ an object dies (a peer leaves)
and with probability  $p_2= \lambda N_c /u$ an object is replaced by two objects (a peer receives the file).

Given $Y_{b}(0)=k$ the extinction time $T_k$ is defined by
\[
%\label{extinction-time}
T_b(k) =\min\{t>0\,:\, Y_{b}(t)=0\}
\]
Let $G_k(t):=P(T_b(k)<t)$ be the CDF of $T_b(k)$. Given $Y_{b}(0)=k$, the extinction probability, $q_k$, is given $q_k=G_k(\infty)$.
The CDF of $T_b(1)$ is obtained from \cite[Eq. (7.3), p. 104]{Ha} and is given by
\begin{equation}
\label{extinction-time-cdf-1}
G_1(t)=\frac{1-e^{-\mu(1-\rho)t}}{1-\rho e^{-\mu(1-\rho)t}}, \quad t\geq 0,
\end{equation}
where we recall that $\rho=\lambda N_c/\mu$. From (\ref{extinction-time-cdf-1}) we find
\begin{equation}
\label{prob-extinction}
q_1 =\min\{1, 1/\rho\}.
\end{equation}
In other words, the extinction will be certain iff $\rho\leq 1$.
Since all objects behave independently of each other we have $q_k=q_1^k=\min\{1,1/\rho^k\}$ and
\begin{equation}
\label{extinction-time-cdf-k}
G_k(t)=G_1(t)^k=\left(\frac{1-e^{-\mu(1-\rho)t}}{1-\rho e^{-\mu(1-\rho)t}}\right)^k, \quad t\geq 0.
\end{equation}
In particular, if $\rho=1$,
\begin{equation}
\label{extinction-time-cdf-k-rho=1}
G_k(t)=\left(\frac{\mu t}{1+\mu t}\right)^k, \quad t\geq 0.
\end{equation}

\subsection{Expected time to extinction}

Assume that $\rho<1$ (extinction is certain).  The expected extinction time
is equal to $E[T_b(k)]=\int_0^\infty (1-G_k(t)) dt$. In particular
\begin{equation}
E[T_b(1)] = -\frac{\log(1-\rho)}{\mu \rho}.
\label{expected-extinction-time}
\end{equation}
Let us now come back to the original process ${\bf Y}$.
Define $T(y_0):=\inf\{t : Y(t)=0\}$,
the first time when the file has disappeared from the network given that $Y(0)=y_0$.
When $Y(0)=Y_b(0)=y_0$, Proposition \ref{prop:bound} implies that
\[
P(T(y_0)>t)=P(Y(t)>0)\leq P(Y_b(t)>0)=G_{y_0}(t).
\]
In particular $E[T(y_0)] \leq E[T_b(y_0)]$, so that
$E[T(1)]\leq  -\frac{\log(1-\rho)}{\mu \rho}$  from (\ref{expected-extinction-time}) for $\rho< 1$ .

\section{Mean-field approximation}
\label{sec:fluid}

In this section we investigate the behavior of the process ${\bf Y}$ defined in Section \ref{sub-general-network}
as $N$, the number of peers, gets large.
We first show that this behavior (to be made more precise) is well approximated by a
deterministic limit solution of an ODE, an approach known as mean-field approximation. See \cite{Kurtz} for the theory
and \cite{Chaintreau, Sigmetrics05:Massoulie, Umass-epidemics} for recent applications
in the area of file sharing systems.

Like in Section \ref{sec:fullycoop} we assume that the pairwise contact rate, $\lambda$, is of the form
$\lambda=\beta/N$ with $\beta>0$. We recall that the initial state of ${\bf Y}$ is given by
\be
\label{init-cond}
Y(0)=N y_0, \quad X_c(0)=N x_{c,0}, \quad X_f(0)=N x_{f,0}
\ee
%with $y$, $x_c$ and $x_f$ arbitrary nonnegative constants such that $y_0+x_{c,0}+x_{f,0}=1$.
with $y_0+x_{c,0}+x_{f,0}=1$.
[The analysis below holds under the weaker condition $\lim_{N\to \infty}N^{-1} (Y(0),X_c(0),X_f(0))=(y_0,x_{c,0},x_{f,0})$.]

Let $v_1=(1,-1,0)$, $v_2=(-1,0,0)$ and $v_3=(0,0,-1)$.
Denote by $g(Y,Y+v_i)$, $i=1,2,3$, the non-zero transition rates of the process Markov process ${\bf Y}$ out of state
$Y=(Y_1,Y_2,Y_3)$.
We have  (cf. (\ref{e:closed1b1})-(\ref{e:closed1b3}))
{\small
\[
g(Y,Y+v_1)=\frac{\beta}{N} Y_1 Y_2, g(Y,Y+v_2)=\mu Y_1,  g(Y,Y+v_3)=\frac{\beta}{N} Y_1 Y_3
\]
}
which can be rewritten as
\begin{equation}
\label{scale-transitions}
g(Y,Y+v_i)=N f\left(\frac{Y}{N},v_i\right),\quad i=1,2,3
\end{equation}
where $f(u,v_1)=\beta u_1 u_2$, $f(u,v_2)=\mu u_1$ and $f(u,v_3)=\beta u_1 u_3$  for $u=(u_1,u_2,u_3)$.

We may therefore use  Theorem 3.1  in \cite{Kurtz} (it is easily that conditions (3.2)-(3.4) in \cite{Kurtz} are satisfied) to obtain that
the rescaled process
$N^{-1} {\bf Y}$  converges in probability as $N\rightarrow \infty$, uniformly on all {\em finite} intervals $[0,T]$, to the solution $(y,x_c,x_f)$,
$0\leq y, x_c, x_f$, $y+x_c+ x_f\leq 1$, of the system of ODEs
\begin{equation}
\frac {d} {dt}
\begin{pmatrix}
y \\ x_c \\ x_f
\end{pmatrix}
= \begin{pmatrix}
y ( \beta x_c - \mu ) \\ - \beta y x_c \\ - \beta y x_f
\end{pmatrix}
\label{ode}
\end{equation}
with initial condition $(y_0,x_{c,0},x_{f,0})$.

In particular,  for any finite $t$ the solution $y$, $x_c$, $x_f$ of (\ref{ode}) will approximate the fraction of peers
with the file, the fraction of cooperative peers without the file and the fraction of free-riders, respectively, at time $t$.
The accuracy of this approximation will increase with $N$, the total number of peers.

\subsection{Peers that never receive the file: a phase transition}
\label{ssec:A}
The fraction of cooperative peers $x_c$ and the fraction of free-riders $x_f$ that
do not have the file monotonically decrease (this is true also for the original
system) to some limit values. They can continue decreasing until there
are no copies of the file in the system, namely until $y=0$.

The first question we wish to address is whether these limits are close to 0 or
are large. In other words, we wish to know whether all (or almost all)
peers interested in the file are able to obtain it or not. If the answer
is no, then we shall be interested in computing the fraction of peers that
never receive the file.

Let $\theta:=\beta/\mu$. From the first two equations in (\ref{ode}) we obtain $x_c$ as
\be
\label{eq:sec4.1.1}
\frac{dy}{dx_c} = -1 + \frac{1}{\theta x_c}.
\ee
\noindent The solution of this differential equation is
\be
\label{eq:sec4.1.2}
x_c + y = \theta^{-1}\ln x_c + \phi(\theta)
\ee
\noindent where $\phi(\theta):= x_{c,0} + y_0 - \theta^{-1}\ln x_{c,0}$.
Let $y^{max}$ be the maximum ratio of cooperative peers with the file.
According to the first equation in (\ref{ode}), $y^{max}$ is reached when
$x_c = \theta^{-1}$ if $\theta>1$ and is expressed as
\be
\label{eq:sec4.1.5}
y^{max} =-\theta^{-1} (1+\ln \theta) + \phi(\theta).
\ee
When $\theta\leq 1$, $y^{\max}$ is reached when $x_c=x_{c,0}$ (i.e. at time $t=0$).
On the other hand, as $t\to \infty$ $y$ is approaching 0 (since we have assumed that $\mu>0$) so that,
from (\ref{eq:sec4.1.2}), $x_c(\infty)$  satisfies the equation
\be
\label{eq:sec4.1.3}
x_c(\infty) - \theta^{-1}\ln (x_c(\infty)) - \phi(\theta) = 0.
\ee
It is easily seen that this equation has a unique solution in $(0,x_{c,0})$ (note that $x_c(t)\leq x_{c,0}$ for any $t$
since $x_c$ is non-increasing from the second equation in (\ref{ode})).
From (\ref{ode}) we find that $x_f(t)= \frac{x_{f,0}}{x_{c,0}}x_c(t)$ for all $t$.

%In particular,  $x_c(\infty)\in (0,\theta^{-1})$ if $\theta x_{c,0}>1$.
As recalled earlier the  mean-field approximation only holds for finite $t$ and there is therefore no guarantee
that it will hold when $t=\infty$, namely, that $N {\bf Y}^{-1}$ will converge in probability to $(0, x_c(\infty),x_{f,0}x_c(\infty)/x_{c,0})$
as $N\to \infty$. However, due to the particular structure of the infinitesimal generator of ${\bf Y}$ this
convergence takes place as shown in  \cite[Sec. 5.2]{darling} (Hint: consider the rescaled Markov process $\tilde {\bf Y}:=
\{(\tilde Y(t), \tilde X_c(t), \tilde X_f(t)), t\geq 0\}$
with generator $\tilde g(\cdot,\cdot)= g(\cdot,\cdot)/Y_1$ and same state-space as ${\bf Y}$, so that starting from the same initial condition  the terminal values of $X_c(t)$ and $ \tilde X_c(t)$ (resp.  $X_f(t)$ and  $\tilde X_f(t)$) will have the same distribution.
The mean-field approximation for $\tilde {\bf Y}$ shows that the solution of the
associated ODE's is given by $(0,x_{c,0}e^{-\beta \tau}, x_{f,0}e^{-\beta \tau})$ for any $t\geq \tau$,
with $\tau$ the unique solution in $(0,\infty)$ of $x_{c,0}+y_0=x_{c,0}\, e^{-\beta \tau}+\mu\tau$, from which the result follows).

In summary, as $N$ is large, the fraction of cooperative (resp. free riders) peers which will never receive the file is approximated
by $x^{\min}_c:=x_{c}(\infty)$ (resp. $x^{\min}_f:=x_{f,0} x^{\min}_c/x_{c,0}$) where  $x_c(\infty)$ can be (numerically) calculated from
(\ref{eq:sec4.1.3}).

We are interested in whether there is an abrupt change in content availability (i.e. $x_c(\infty)$) with the parameter $\theta$.
Obviously, if $\theta$ is 0, all the cooperative peers that do not have the file at time 0 will never receive it.
To find a phase transition, we approximate $\log(x_c(\infty))$ in (\ref{eq:sec4.1.3}) by using its Taylor extension at $x_{c,0}$
and obtain
\be
\label{eq:sec4.1.4}
x_c(\infty) \approx ((\frac{1}{\theta} - x_{c,0}-y_0)+ \frac{1}{\theta}(\frac{x_c(\infty)}{x_{c,0}} -1)^2)\big/(\frac{1}{\theta x_{c,0}}-1). \nonumber
\ee
Since the expression $\frac{1}{\theta}(\frac{x_c(\infty)}{x_{c,0}} -1)^2$ is bounded, the phase transition happens at $\theta  = 1/x_{c,0}$.

Despite the similarity in the  definitions of  $\rho$ in Section \ref{sec:branch} and of $\theta x_{c,0}$ in the present section,
 the phase transition at $\rho=1$ is different in nature from that  at $\theta x_{c,0}=1$.  The former indicates whether or
not the file will be extinct while the latter will drastically impact the final size of the torrent.

\begin{figure}[ht]
%\begin{minipage}[ht]{50mm}
\begin{center}
\includegraphics[height = 1.4in,width=2.7in]{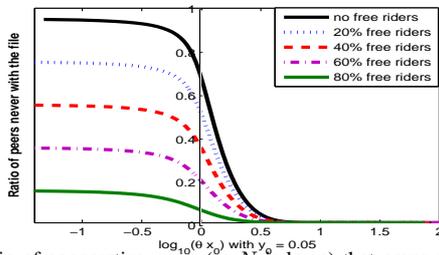}
\end{center}
%\end{minipage}
\vspace{-0.7cm}
\caption{\label{fig1}
Ratio of cooperative peers (as $N$ is large) that never receive the file  as a function of
$\log_{10}(\theta x_{c,0})$.}
\end{figure}

Figure \ref{fig1} displays the mapping $\log_{10}(\theta x_{c,0})\to x_c^{\min}$
for $x_{c,0}\in\{0.01, 0.1, 0.3, 0.5,0.9\}$ and $y_0=0.05$.
The curves for $x_c^{\min}$ are monotonically decreasing in $x_{c,0}$
(the curve that intersects the vertical axis close to $1$ is
the one corresponding to $x_{c,0}=0.01$, and so on.).
For each curve we note the existence of a phase transition at $\theta x_{c,0}=1$, which is more pronounced
as the ratio of free riders increases.
\vspace{-0.2cm}
\vspace{-0.3cm}

\begin{comment}
We see from Fig. \ref{fig1} (right) a clear phase transition
that occurs at $\theta=1$ (i.e. when $\beta=\mu$).
$x_c^{\min}$ is close to $x_{c,0}$ (see (\ref{eq:sec4.1.3}))
that depends on the initial condition) as $\theta$ decreases to zero,
and approaches zero as $\theta $ increases beyond $1$ (or equivalently
as $\lambda$ increases beyond $\mu$).
\end{comment}

\subsection{Combining the branching and the epidemic model}
\label{ssec:B}
The mean-field approximation is accurate for large $N$ if the initial state scales
with $N$ linearly. In the case
that $N$ is very large but the initial condition does not scale with $N$ (e.g. $Y(0)=1$,
$X_c(0)+X_f(0)=N-1$),
we can do the following. Fix some $N_0 $ much smaller than  $N$ but larger than $1$.
Use the branching process approximation until the number of peers with the file is $N_0$.
Then, switch to the epidemic model. (For the branching process, we recall
that given that there is no extinction, the population size grows
exponentially fast). % {\bf to complete}

\section{Control actions against P2P networks}
\label{sec:control}

In this section, we first investigate the major findings in the
analysis of content availability. A set of control actions
are proposed to protect copyrighted files against P2P
file sharing.

\subsection{Observations on file availability}
\label{ssec:file-availability}

Before proposing the counteractions against illegal P2P swarms,
we investigate the impact of measures on file availability.
The main question is how does a decrease or increase in one of the system parameters
affect measures such as
\begin{itemize}
\item
the size of the torrent: the fraction of those who are
interested in the file and are able to get a copy of it.
This can be seen as a global availability measure.
\item
the extinction probability or the expected extinction time,
\item
the maximum availability: the maximum number of copies that
can be found simultaneously in the system. This can be
viewed as an instantaneous availability measure.
\end{itemize}

According the analysis in Sections \ref{sec:branch}-\ref{sec:fluid}, all above measures depend on
the ratio $\frac{\lambda}{\mu}$ (or $\frac{\beta}{\mu}$ equivalently).
A small ratio $\frac{\lambda}{\mu}$
means a poor availability of the file.
However, the contact rate $\lambda$ is an intrinsic parameter
of P2P swarms that can hardly be changed technically.
An even more challenging problem is that there usually exist
several illegal publishers residing in the system for a very long time.
They aim to spread the copyright protected file as wide as possible in the P2P swarm.
To combat with undesirable file sharing, we present two methods,
the cooperation control and the pollution attack. The former is to discourage the
degree of cooperation of peers with the file. The latter introduces a number
of polluters before the file dissemination begins, which can be found in \cite{TR10}
due to the page limit of this paper.
\subsection{Control of cooperation}
We introduce the cooperation control to prevent the dissemination of copyrighted files.
We aim to reduce the degree of cooperation (i.e. increasing $\mu$) so that the delay
of obtaining the file is increased.
To achieve this goal, the content owner can invest a certain amount of money
in the very beginning to discourage the cooperation of peers.
The cooperation control
does not contradict with our opposition of collaborative solutions such as flat tax.
In fact, we are focusing on this unilateral action of the content owner against unauthorized file dissemination.

We consider the same model as in Section \ref{sub-general-network} but we now assume that
all peers are cooperative and that there is a number $Y_N^*>0$ of permanent publishers, where the subscript $N$
refers to the total number of peers in the system at time $t=0$. The pairwise contact rate is
$\lambda=\beta/N$.
Denote by $\alpha$ the investment level of the content owner against
P2P networks. The departure rate is an increasing function of $\alpha$, denoted by $\mu(\alpha)$.
We denote
by $Y_N(t)$ the number of non-permanent publishers and by $X_N(t)$ the number of peers without the file at time $t$.
Observe that  $Y_N(0)+X_N(0)=N-Y^*_N$.
If $\lim_{N\to \infty} Y_N(0)/N=y_0$ and  $\lim_{N\to \infty} X_N(0)/N=x_0$, which implies  that
$\lim_{N\to \infty} Y^*/N=1-x_0-y_0:=y^*$,  then, by Kurtz's result \cite{Kurtz}, the rescaled process $N^{-1}(Y(t),X(t))$
converges in probability as $N\to \infty$, uniformy on all finite intervals $[0,T]$, to the solution of the ODEs
\begin{equation}
\frac {d} {dt}
\begin{pmatrix}
y \\ x
\end{pmatrix}
= \begin{pmatrix}
 \beta (y + y^*)x - \mu(\alpha) y \\ -\beta (y+y^*)x
\end{pmatrix}
\label{ode3}
\end{equation}
\noindent with  initial state $(y_0,x_0)$.  From now on we will assume that $y^*>0$.

Consider an arbitrary peer without the file at time $t=0$ and denote by $T_N$ the
time that elapses before it receives it.  Let  $P_N(t):=P(T_n<t)$.
Similarly to \cite[Page 6]{Umass-epidemics} we find
\be
\frac{dP_N(t)}{dt} = \beta (1-P_N(t))\frac{E[Y_N(t)]+ Y^*_N}{N}.
\ee
Solving for $P_N(t)$ gives
\begin{equation}
\label{ode-sectionV}
P_N(t)=1-e^{-\beta \int_0^t \frac{E[(Y_N(s)]+Y^*_N}{N} ds},\quad t\geq  0.
\end{equation}
Hence
\[
E[T_N]=\int_0^\infty (1-P_N(t))dt = \int_0^\infty e^{-\beta \int_0^t  \frac{E[Y_N(s)]+Y^*_N}{N} ds}dt.
\]
From the above we know that  $(E[Y_N(t)]+Y^*_N)/N\to y+y^*$ as $N\to\infty$ for every $t>0$, so that
from (\ref{ode-sectionV})
\begin{equation}
\lim_{N\to\infty}(1-P_N(t))=e^{-\beta \int_0^t (y(s)+y^*) ds}
\label{bc-1}
\end{equation}
for every $t>0$.
On the other hand, $\lim_{N\to\infty} Y^*_N/N=y^*$ implies that for $0<\epsilon<y^*$ there exists
$N_{\epsilon}$ such that $Y^*_N/N> y^*-\epsilon$ for all $N>N_\epsilon$. Therefore, from (\ref{ode-sectionV}),
\begin{equation}
\label{bc-2}
1-P_N(t) \leq e^{-\beta (y^*-\epsilon) t}
\end{equation}
 for $N>N_\epsilon$, $t >0$.  Since the r.h.s of (\ref{bc-2}) is integrable in $[0,\infty)$,
(\ref{bc-1}) and (\ref{bc-2}) allow us to apply the bounded convergence theorem to conclude that
 \[
 \overline{T}(\alpha):=\int_0^\infty \lim_{N\to \infty} (1-P_N(t))dt=\int_0^\infty e^{-\beta \int_0^t (y(s)+y^*) ds} dt.
 \]
\noindent The objective of the content owner is to choose an investment level $\alpha\geq 0$ which will maximize its utility
\be
\label{utility}
h(\alpha) := \overline{T}(\alpha)- \alpha.
\ee
To understand the impact of cooperation control on
the delay, we present numerical studies in Section \ref{sec:impact}.
%The utility of content owner is satisfactory
%when the investment level makes ratio $\beta/\mu(\alpha)$ two or three times larger than the threshold  one.

%{\bf A revoir: definition of $\rho$?}

\begin{comment}
\subsection{More on the mean-field}

In the above optimization problem, we could control
several parameters, such as $\beta$ and $\mu$.
For each parameter, we know by Kurtz's theorem that the
discrete value converges to the mean-field value. This does
not imply in general that the optimal value and optimal
parameter in the mean-field limit are limits of the optimal
value and parameter in the initial discrete models.

If the set of parameters which we denote by $G$,
is convex and compact, and if the
costs of the $N$th model converge to that under
the mean-field limit (as $N \to \infty$) {\it uniformly}
over the set $G$, then indeed the optimal value of
the discrete models converges to the value of the
mean-field  optimization problem. Moreover, any limit of a
subsequence of parameters that are optimal for the discrete
models is optimal for the limit problem.

This statement remains true even when we allow the parameters
to depend on the state (feedback controls), as long as these
controls are within a class of controls that has a uniform
Lipschitz control.
\end{comment}

\section{P2P with a fixed request rate per node}
\label{sec:modelSi}

\subsection{Model}
In this section we will consider a slight variation of the model in \cite{Si}: there are
$N$ peers at time $t=0$, at least one of them having a file.
Each peer without the file sends a request for the file
to another peer selected at random.  These requests are initiated at Poisson rate
$\lambda>0$. It is assumed that a peer with the file leaves the system after an exponentially
distributed random duration with rate $\mu>0$. All these rvs are mutually independent.
Let $Y(t)$ (resp. $X(t)$) be the number of peers with the file (resp. without the file)
at time $t$. We have $Y(0)+X(0)=N$ with $Y(0)\geq 1$. Under the above assumptions ${\bf Z}:=\{(Y(t),X(t)),t\geq 0\}$ is
a Markov process on the set ${\cal E}:=\{(y,x)\in \{0,1,\ldots,N\}^2\,: 0\leq y+x\leq N\}$. Let
$q(z,z')$, $z=(y,x), z'=(y,x)\in {\cal E}$, denote its generator. Non-zero transition rates are given by
\bel{eq:closed1c1}
\begin{pmatrix}
Y (t) \\ X(t)
\end{pmatrix}
\to
\begin{pmatrix}
Y (t)+1 \\ X (t)- 1
\end{pmatrix}
\mbox{ with rate $\displaystyle \frac{ \lambda Y(t) X(t)  }{ Y(t)+X(t)}$,}
\ee
\bel{eq:closed1c2}
\begin{pmatrix}
Y (t) \\ X(t)
\end{pmatrix}
\to
\begin{pmatrix}
Y (t) -1 \\ X(t)
\end{pmatrix}
\mbox{ with rate $\mu Y(t) $.}
\ee
This model differs from our previous model in that the rate of increase is normalized by the total number of peers in
the system. More precisely, the rate in (\ref{eq:closed1c1}) follows from the fact that with probability $Y(t)/(Y(t)+X(t))$ a peer
without the file will contact a peer with a file at time $t$ (the latter implicitly assumes that a peer may contact itself
as otherwise this probability  would be $Y(t)/(Y(t)+X(t)-1)$; the reason for doing this  will next become apparent.
Note that this assumption will have no effect when $N$ gets large) so that the total rate of increase of the number
of peers with the file is  $\lambda Y(t) X(t)/(Y(t)+X(t))$ at time $t$.

The same model is considered in \cite{Si} with the difference that  in \cite{Si} there is one
permanent publisher, thereby implying that all peers will receive the file. These authors show that
that the mean broacast time is $O(N)$ if $\lambda<\mu$ and is
$O(\log(N))$ if $\lambda>\mu$.  Thus, there is a phase transition at $\lambda=\mu$.

In this section we will instead focus on (i) the fraction of peers that will receive the file (in the absence of a permanent publisher
this fraction is not always equal to 1)
and on (ii) the maximum torrent size (maximum number of copies of the file at one time) as $N$ is large.
In both cases we will show the existence of phase transitions.

Our analysis will use Kurtz's theorem \cite[Thm 3.1]{Kurtz} like in Section \ref{sec:fluid}. Note, however, that both metrics (i) and (ii)
above require to use the mean-field limit as $t\to\infty$, something that Kurtz's result does not cover.

To overcome this difficulty, we will use the same argument as in \cite{darling} (see also Section \ref{sec:fluid} where this argument
was already used), taking advantage of the particular structure of the infinitesimal generator of the process ${\bf Z}$.
More specifically,  it is seen that the generator of ${\bf Z}$  writes in the form $q(z,z')=
y \tilde q(z,z')$ for $z=(y,x),z'=(y',x')\in {\cal E}$, where non-zero transition rates are given in (\ref{eq:closed1c1})-(\ref{eq:closed1c2}).

Let ${\bf \tilde Z}=\{(\tilde Y(t),\tilde X(t)),t\geq 0\}$ be a Markov process with generator $\tilde g(z,z')$ and state-space ${\cal E}$ (same state-space
as ${\bf Z}$).

Since ${\bf \tilde Z}$ has been obtained by changing the time-scale of ${\bf Z}$, the final values of $\tilde X(t)$ and of $X(t)$ will have
the same distribution (note that the final state of $\tilde Y(t)$ and $Y(t)$ is always zero since states $(0,\cdot)$ are all absorbing states)
and so will have the maximum torrent size.

Since the generator $\tilde g(z,z')$ can be written as $\tilde g(z,z')=N f(z/N,z'/N)$ (this is where the assumption that a peer may contact itself is useful)
and since conditions (3.2)-(3.4) in \cite[Thm 3.1]{Kurtz} are clearly  satisfied,
we may apply \cite[Thm 3.1]{Kurtz} to obtain that, at any finite time $t$,
 $N^{-1}(\tilde Y(t),\tilde X(t))$ converges in probability as $N\to \infty$ to  the solution $(\tilde y, \tilde x)$, $0\leq \tilde y,\tilde x\leq 1$,
$\tilde y +\tilde x \leq 1$, of the ODEs
\begin{equation}
\label{ode-sindo-scaled}
\dot {\tilde y} = -\mu + \lambda\tilde x/(\tilde x+\tilde y),\quad \dot{\tilde x}= -\lambda\tilde x/(\tilde x+\tilde y)
\end{equation}
given that $\lim_{N\to\infty} N^{-1} (\tilde Y(0),\tilde X(0))=(\tilde y(0),\tilde x(0))$.  Let
$(y_0,x_0):=(\tilde y(0),\tilde x(0))$. We will assume that $0<y_0<1$ and $y_0+x_0=1$ (the case $y_0=0$ (resp. $y_0=1$) has no interest since it corresponds
to a P2P network with no file at any time (resp. where all peers have the file at time $t=0$).

\subsection{Phase transitions}

Define $\xi:=\lambda/\mu$.
Adding both ODEs in (\ref{ode-sindo-scaled}) yields $\tilde y(t)+\tilde x(t)=-\mu t +1$. Plugging this value back into
(\ref{ode-sindo-scaled}) gives $\tilde x(t)=x_0 (1-\mu t)^{\xi}$ for $0\leq t <1/\mu$ and, by continuity,
$\tilde x(t)=x_0 (1-\mu t)^{\xi}$ for $0\leq t \leq 1/\mu$ with $\tilde x(1/\mu)=0$.

In order to approximate the fraction of peers which will never receive the file as $N$ is large, one needs to find
the first time $\tau>0$ where either $\tilde x(\tau)=0$ or $\tilde y(\tau)=0$. This time  $\tau$ is easy to find as shown below.

We already know that $\tilde x(t)>0$ for $0\leq t<1/\mu$ and $\tilde x(1/\mu)=0$ so that we only need to focus on the zeros
of $\tilde y(t)$ in $[0,1/\mu]$.
By writing $\tilde y$ as $\tilde y(t)=(1-\mu t)(1-x_0 (1-\mu t)^{\xi-1})$ we conclude that the smallest zero of $\tilde y$ in
$[0,1/\mu]$ is $(1-x_0^{1/(1-\xi)})/\mu$ if $\xi<1 $  and is $1/\mu$ if $\xi\geq 1$. Therefore,
$\tau=(1-x_0^{1/(1-\xi)})/\mu>0$ if $\xi<1$  and $\tau=1/\mu$ if $\xi\geq 1$.
Introducing this value of $\tau$ in $\tilde x(t)$ yields $\tilde x(\tau)=x_0^{1/(1-\xi)}$ if $\xi<1$ and
$\tilde x(\tau)=0$ if $\xi\geq 1$.
In other words, as $N$ is large, all peers will get the file if $\xi\geq 1$ and a fraction $x_0^{1/(1-\xi)}$ of them will not if $\xi<1$.
In other word we observe a phase transition at $\xi=1$: all peers will get the file if $\xi\geq 1$ and a fraction
 $\xi\to x_0^{1/(1-\xi)}$ will not if $\xi<1$.

\begin{figure}[ht]
%\begin{minipage}[ht]{50mm}
\begin{center}
\includegraphics[height = 1.4in,width=2.7in]{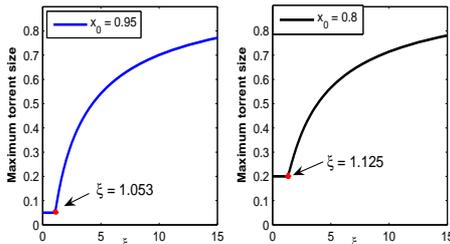}
\end{center}
%\end{minipage}
\vspace{-0.7cm}
\caption{\label{fig1-bis}
Maximum torrent size over $N$ (as $N$ is large) as a function of $\xi$ for
$x_0=0.95$ (left figure) and $x_0=0.8$ (right figure).}
\end{figure}

Let us now turn to the maximum torrent size. As $N$ is large it will be approximated by the maximum of $\tilde y$ over the interval $[0,\tau]$.
A straightforward analysis of the mapping $t\to \tilde y(t)$ in $[0,\tau]$ shows that
\begin{itemize}
\item
it is decreasing if $\xi\leq 1$ or if $\xi>1$ and $\xi x_0\leq 1$ -- these conditions can be merged into the single condition $\xi \leq 1/x_0$ --
so that its
maximum, $y^{\max}$, is given by $y^{\max} =y_0=1-x_0$,
\item
it is unimodular (first increasing then decreasing) if $\xi>1/x_0$,
with its maximum reached at $t_1:=(1-(\xi x_0)^{1-\xi})/\mu$ and given by
\begin{equation}
\label{max-torrent}
y^{\max}= (x_{0})^{1/(1-\xi)} \xi^{\xi/(1-\xi)} (\xi- 1)>0.
\end{equation}
\end{itemize}
In summary,  as $N$ is large, the maximum torrent size is approximated by $N y^{\max}$ with $y^{\max}$ given in (\ref{max-torrent}) if $\xi >1/x_0$ and $y^{\max}=1-x_0$ if $\xi \leq 1/x_0$. This shows another phase transition (see Fig. \ref{fig1-bis}) at $\xi=1/x_0$ (i.e. at $\lambda x_0=\mu$) in the sense that the torrent is maximum at $t=0$ if   $\xi \leq 1/x_0$  and is maximum at a later time  if $\xi >1/x_0$.
\vspace{-0.2cm}
\vspace{-0.3cm}

%\section{The impact of measures against P2P networks}
\section{Numerical results}
\label{sec:impact}

This section has two goals: to investigate the accuracy of the approximations developed
in the previous sections  (to be made more precise) and to study the impact of measures
against non-authorized uploading or downloading on the file availability in P2P swarming systems.
 Due to lack of space, we will not report any
numerical result for the P2P model considered in Section \ref{sec:modelSi}; we will instead focus on the P2P
model introduced in Section \ref{sub-general-network} and on its branching and mean-field approximations
developed in Sections \ref{sec:branch} and \ref{sec:fluid}, respectively (Fig. \ref{fig:lambda_exttimecdf_no3}-\ref{fig:mu_peernofile}),
as well as on the optimization problem set in Section \ref{sec:control} (Fig. \ref{fig:control1}).

For each set of parameters, between 200 and 1000 discrete-event simulations of the Markov model in Section \ref{sub-general-network}
have been run. In each figure (except in Fig. \ref{fig:mu_exttimecdf_no1}-\ref{fig:mu_exttimecdf_no2} where only simulation results
are displayed and in Fig. \ref{fig:control1} where only mean-field results are shown)  both simulation and approximation results are reported for
the sake of comparison. Let  $r:=(Y(0)+X_c(0))/N$ be the ratio of cooperative peers at time $t=0$ and recall that $N_c=X_c(0)$ (see Section \ref{sec:model}).
The total number of peers, $N$,  at time $t=0$ is equal to $400$ in Fig. \ref{fig:lambda_exttimecdf_no3}-\ref{fig:mu_exttimecdf_no2},   to $300$ in
Fig. \ref{fig:lambda_exttime_no1} -\ref{fig:lambda_exttime_no2} and to $500$ in Fig. \ref{fig:control1}.

\subsection{File extinction time and the branching approximation}

\begin{figure*}[!htb]
 \begin{minipage}{0.3\linewidth}
    \centering
   \includegraphics[width=2.4in, height = 1.9in]{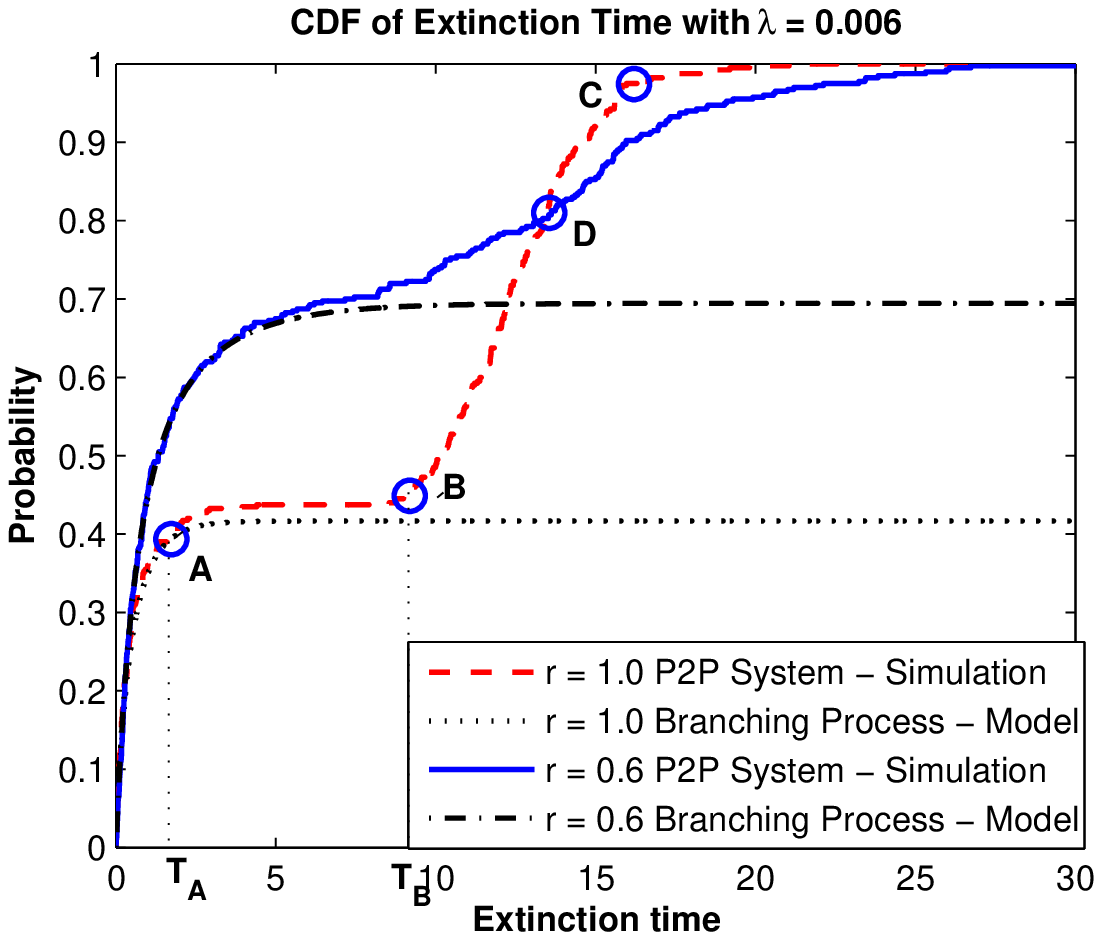}
   \caption{\small{CDF of extinction time for  $Y(0) = 1$, $\lambda = 0.006$, $\mu=1$.}}
   \label{fig:lambda_exttimecdf_no3}
 \end{minipage}
 \hspace{0.5cm}
 \begin{minipage}{0.3\linewidth}
    \centering
   \includegraphics[width=2.4in, height = 1.9in]{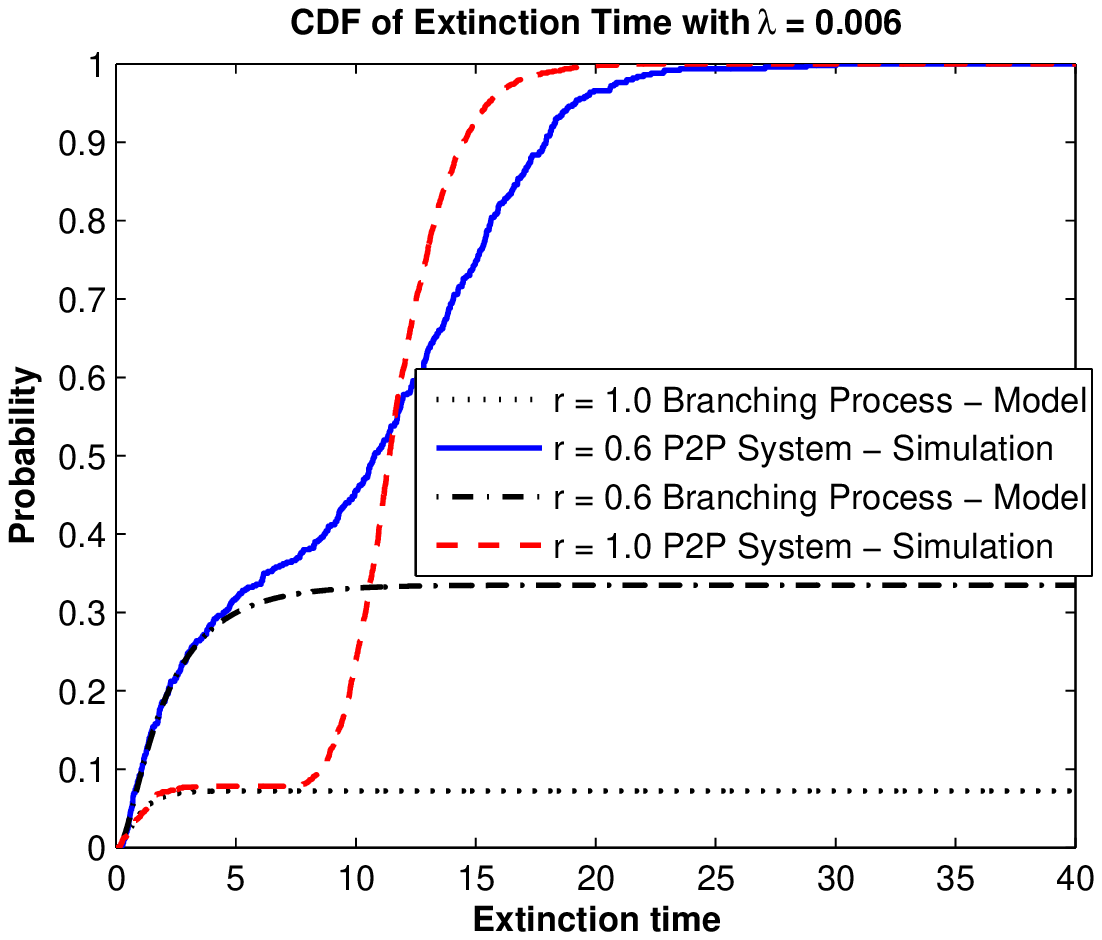}
   \caption{\small{CDF of extinction time for $Y(0) = 3$, $\lambda = 0.006$, $\mu=1$.}}
   \label{fig:lambda_exttimecdf_no4}
 \end{minipage}
 \hspace{0.5cm}
 \begin{minipage}{0.3\linewidth}
   \centering
   \includegraphics[width=2.4in, height = 1.8in]{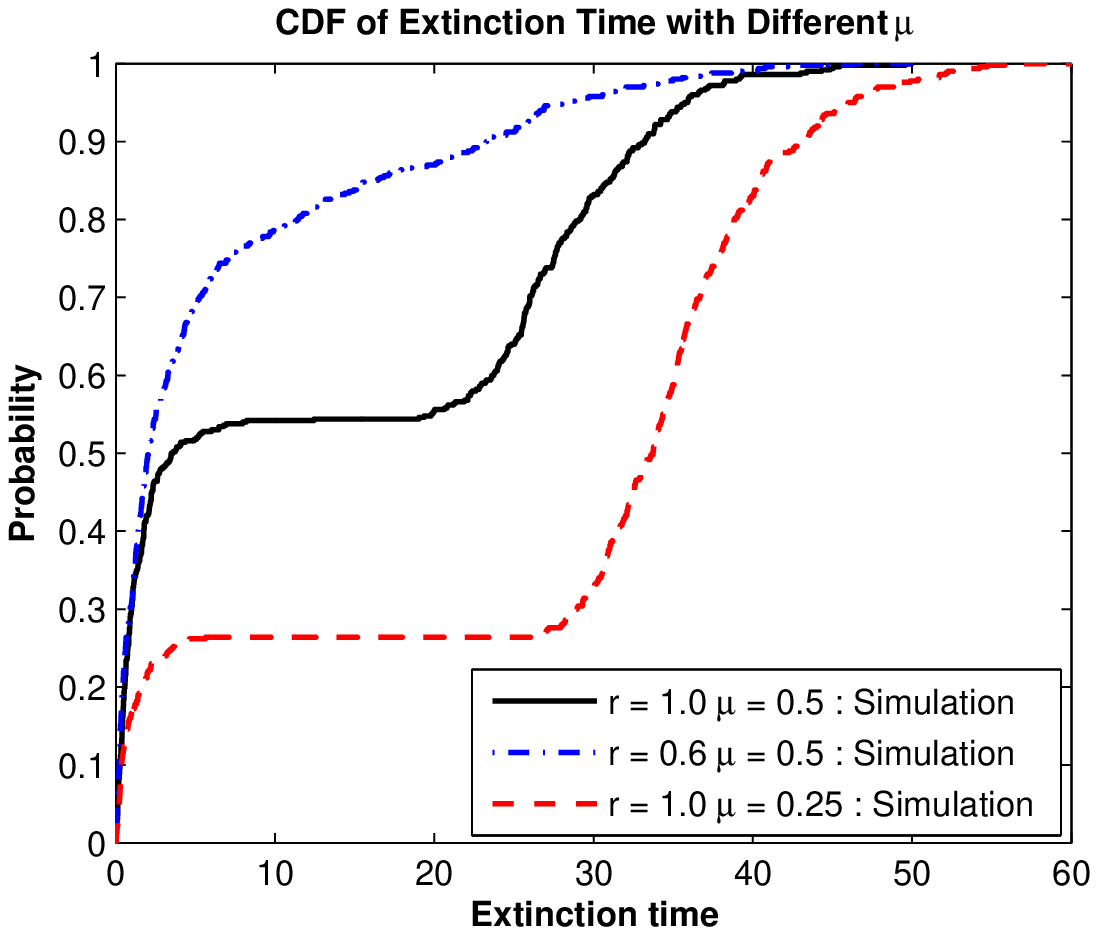}
   \caption{\small{CDF of extinction time  for $Y(0) = 1$ and different $\mu$.}}
   \label{fig:mu_exttimecdf_no1}
 \end{minipage}
\vspace{-0.3cm}
\end{figure*}

\begin{figure*}[!htb]
 \begin{minipage}[t]{0.3\linewidth}
   \centering
   \includegraphics[width=2.4in, height = 1.8in]{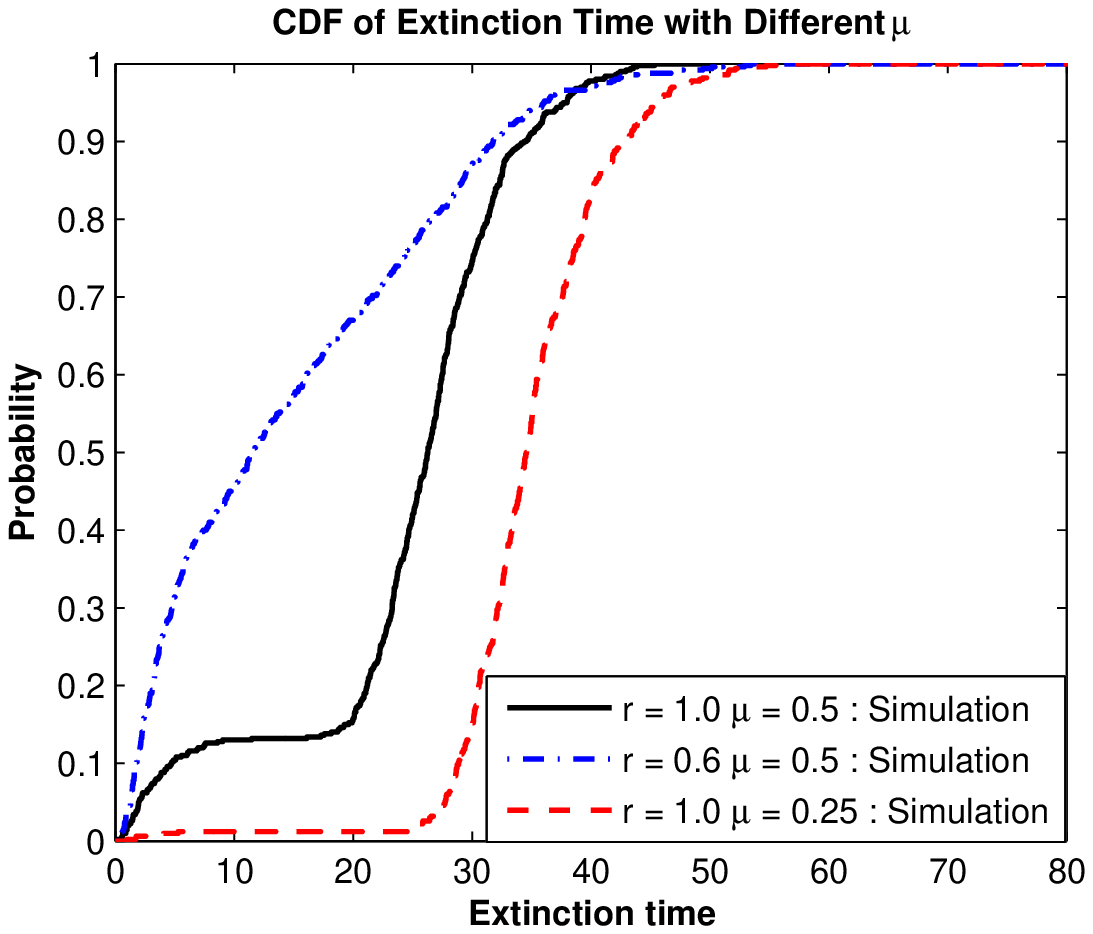}
   \caption{\small{CDF of extinction time for $Y(0) = 3$ and different $\mu$.}}
   \label{fig:mu_exttimecdf_no2}
 \end{minipage}
 \hspace{0.5cm}
 \begin{minipage}[t]{0.3\linewidth}
   \centering
   \includegraphics[width=2.4in, height = 1.8in]{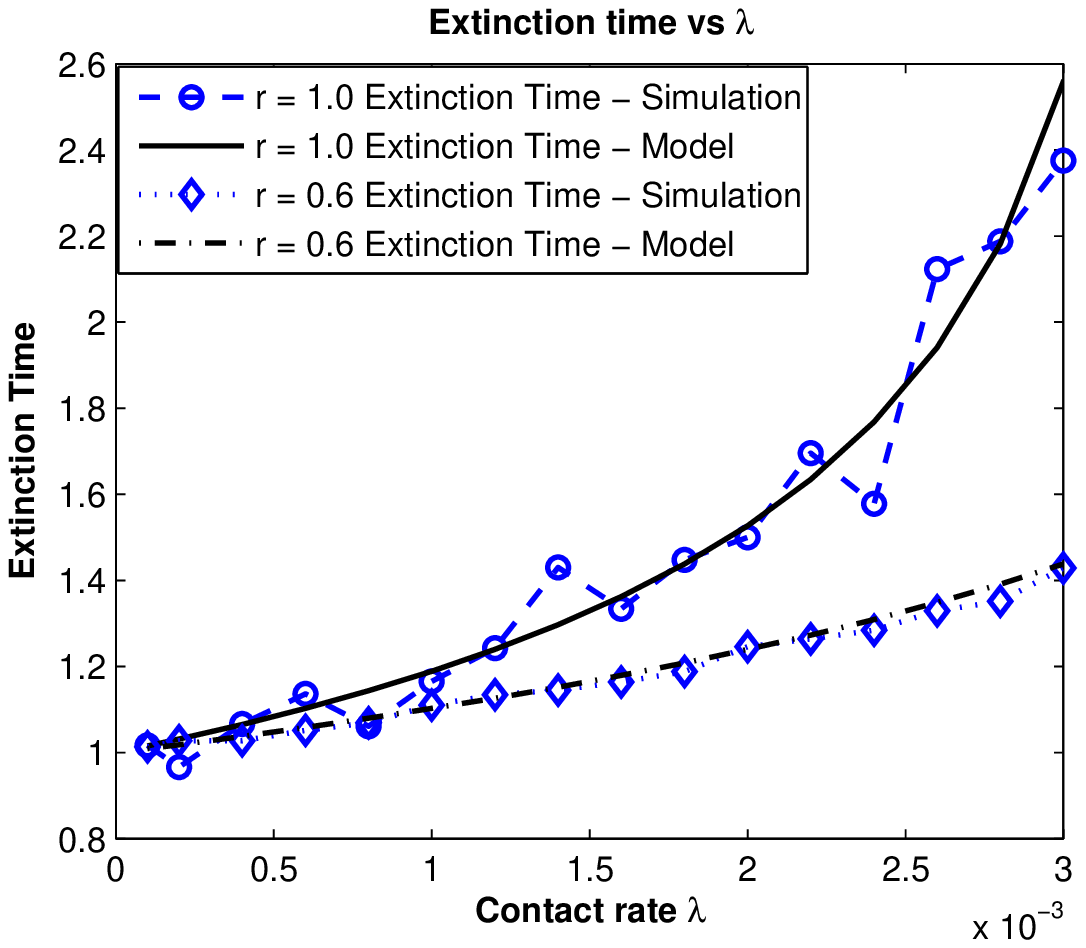}
   \caption{\small{Early extinction time as a function of  $\lambda$ with $Y(0) = 1$.}}
   \label{fig:lambda_exttime_no1}
 \end{minipage}
 \hspace{0.5cm}
 \begin{minipage}[t]{0.3\linewidth}
   \centering
   \includegraphics[width=2.4in, height = 1.8in]{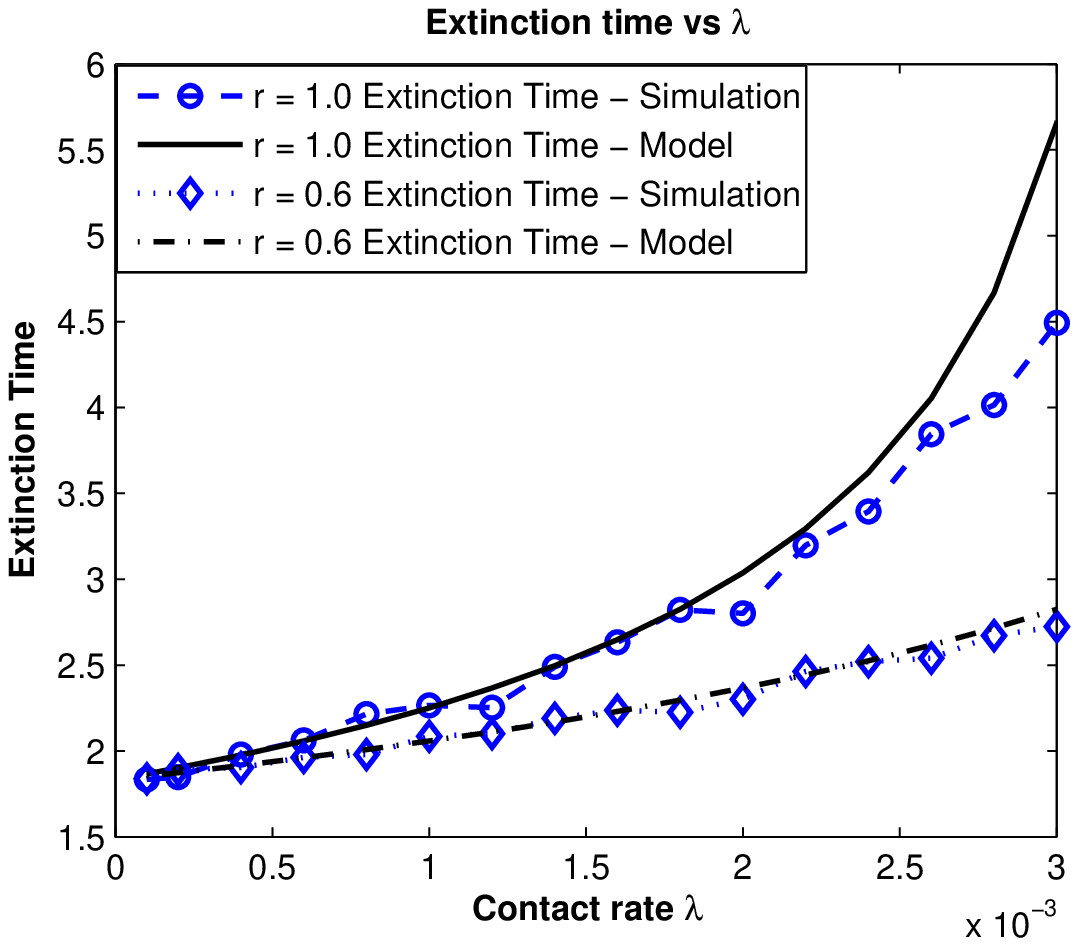}
   \caption{\small{Early extinction time as a function of $\lambda$ with $Y(0) = 3$.}}
   \label{fig:lambda_exttime_no2}
 \end{minipage}
\vspace{-0.3cm}
\end{figure*}

In this section we focus on Fig. \ref{fig:lambda_exttimecdf_no3}-\ref{fig:lambda_exttime_no2}.
Fig. \ref{fig:lambda_exttimecdf_no3} (resp. Fig \ref{fig:lambda_exttimecdf_no4}) compares the CDF of the extinction time obtained by simulation and by the
branching approximation in (\ref{extinction-time-cdf-k}) when $Y(0)=1$ (resp. $Y(0)=3$), $\lambda=6 \cdot 10^{-3}$, $\mu=1$,
and for two values of $r$ ($r=0.6$ implies that $N_c=239$ and $X_f(0)=160$,  $r=1$ implies that there are no free riders ($X_f(0)=0$) and $N_c=399$).
Note that $\rho=\lambda N_c/\mu$ is close to $2.4$ when $r=1$ and is close to $1.43$ when $r=0.6$.
In all cases, the simulation and the branching approximation are in close agreement up to a certain time (time $T_B$ in Fig.  \ref{fig:lambda_exttimecdf_no3})
which, interestingly, corresponds to the extinction time in the branching model. After this time, the extinction of the file in the Markov model increases
sharply (the larger $r$ the larger the increase). In other words, the extinction of the file in the original Markov ml has two modes,
an {\em early} extinction mode and a {\em late} extinction mode. The former occurs when the file disappears before the dissemination has reached its peak
value (i.e. most peers do not get the file) and the latter when most peers leave the network with the file. One may also check that the  branching approximation
provides  an upper bound  for the CDF of the extinction time, as predicted by Proposition \ref{prop:bound}. Last, we note that when there are less cooperative peers ($r=0.6$) the file lifetime is prolonged (see e.g. point $D$ in Fig. \ref{fig:lambda_exttimecdf_no3} where simulation curves for $r=1$ and $r=0.6$ cross each other); this can be explained by the fact that there are less contact opportunities between cooperative peers when $r=0.6$.
The main difference between Fig. \ref{fig:lambda_exttimecdf_no3} and Fig \ref{fig:lambda_exttimecdf_no4} lies in the
increase of the probability of the late extinction that is steeper with three initial seeds ($Y(0)=3$) than with one initial seed ($Y(0)=1$).

Simulation results in Fig. \ref{fig:mu_exttimecdf_no1}-\ref{fig:mu_exttimecdf_no2} exhibit the same {\em early-late} extinction pattern as in
Fig. \ref{fig:lambda_exttimecdf_no3}- \ref{fig:lambda_exttimecdf_no4}; they have been obtained for $\lambda=25\cdot 10^{-4}$ and
for two different values of  $\mu$, $r$ and $Y(0)$.

Fig. \ref{fig:lambda_exttime_no1}-\ref{fig:lambda_exttime_no2} show the expected time to extinction as a function of the pairwise contact rate $\lambda$,
in the case of an early extinction (i.e. for small values of $\lambda$), for  $\mu=1$ and for two values of $r$.
The curves "Model" display the mapping $\lambda \to E[T_b(k)]$, with  $E[T_b(k)]$
the expected extinction time in the branching process  given $Y(0)=k$ (see Section \ref{sec:branch}).
We observe an excellent match between the simulation and the branching approximation thereby showing that the latter works well for early file extinction.
Also note that having three seeds instead of one greatly extends the expected extinction time.

\subsection{File availability and the mean-field approximation}

\begin{figure*}[!htb]
 \begin{minipage}[t]{0.3\linewidth}
   \centering
   \includegraphics[width=2.5in, height = 1.8in]{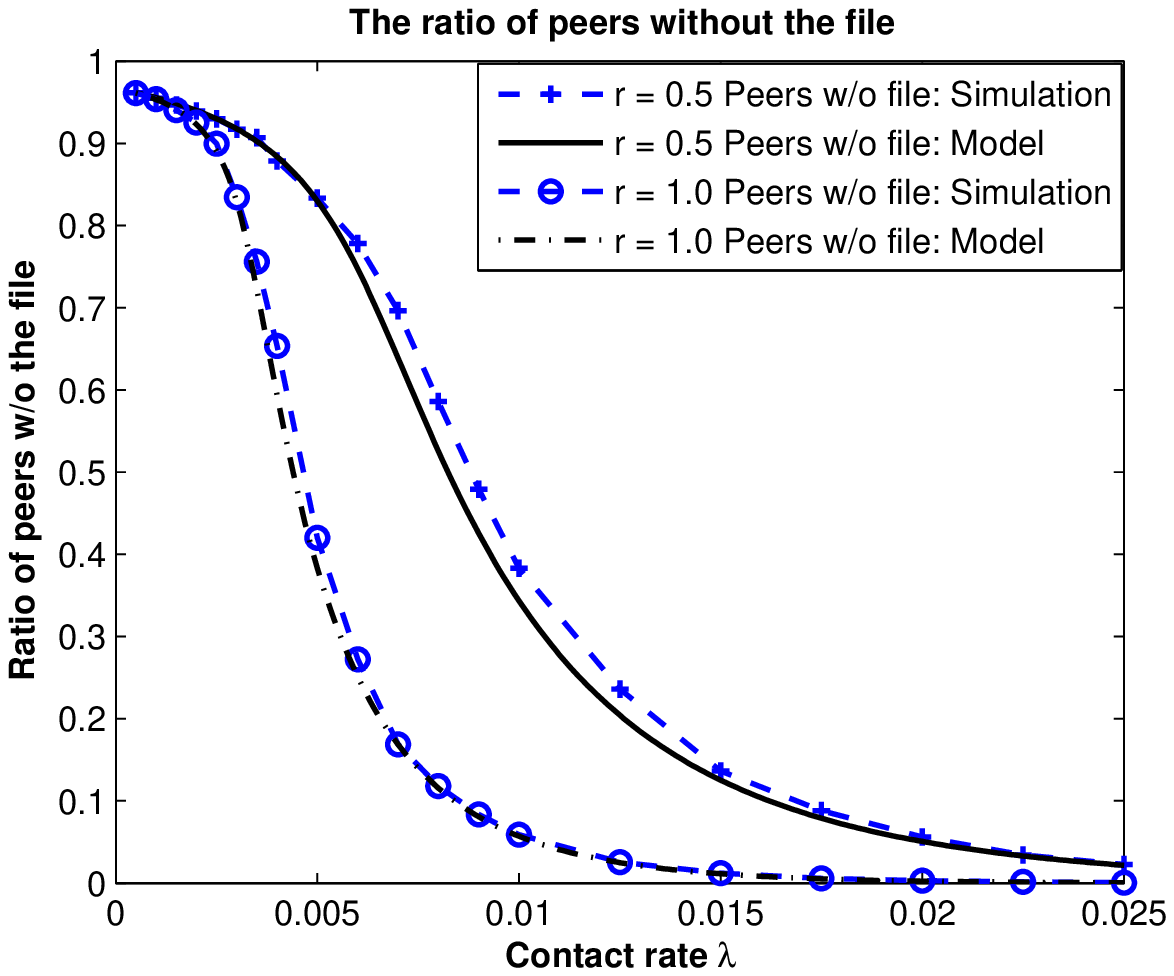}
   \caption{\small{Fraction of peers without the file as a function of  $\lambda$.}}
   \label{fig:lambda_nofile}
 \end{minipage}
 \hspace{0.5cm}
 \begin{minipage}[t]{0.3\linewidth}
   \centering
   \includegraphics[width=2.4in, height = 1.7in]{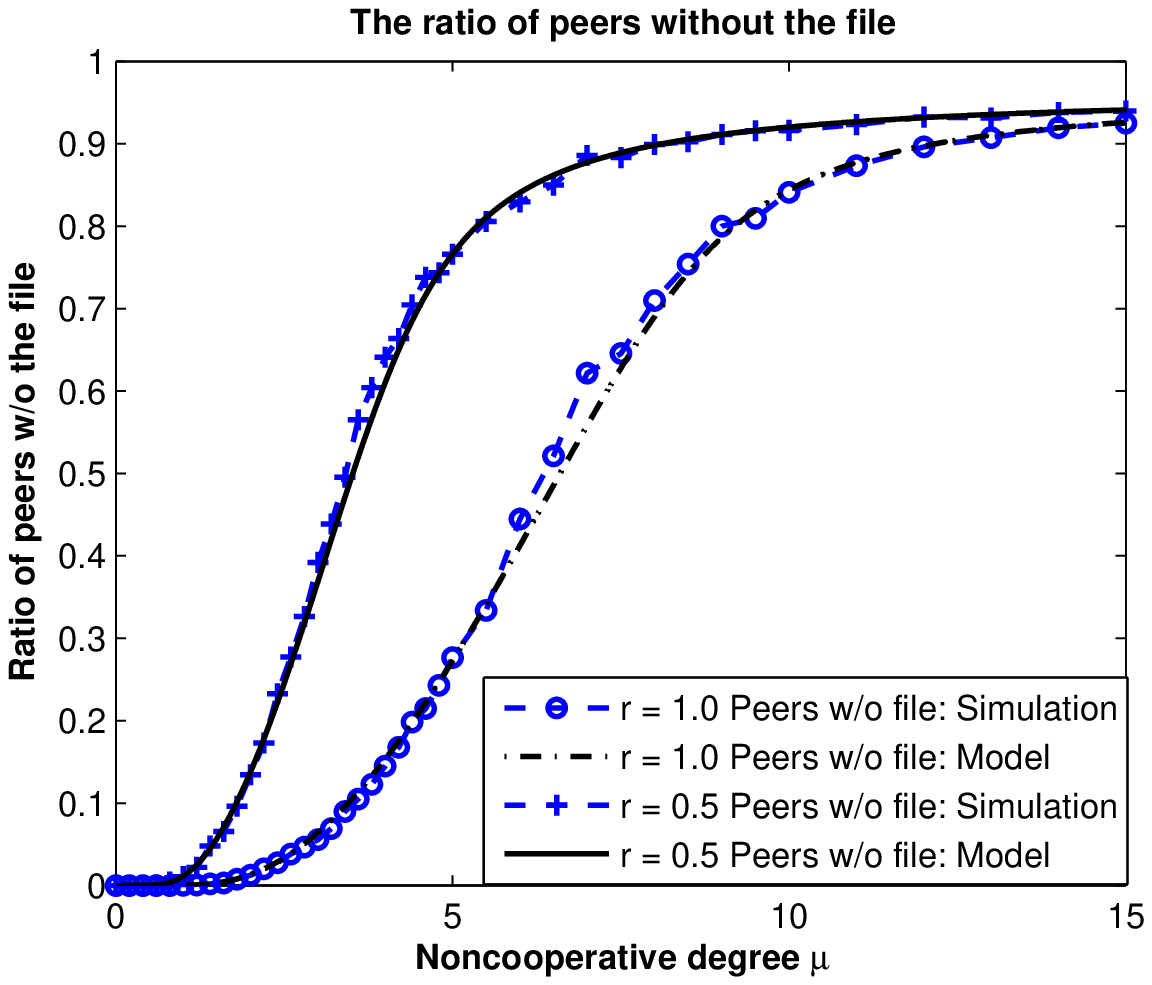}
   \caption{\small{Fraction of peers without the file as a function of  $\mu$.}}
   \label{fig:mu_peernofile}
 \end{minipage}
 \hspace{0.5cm}
 \begin{minipage}[t]{0.3\linewidth}
   \centering
   \includegraphics[width=2.4in, height = 1.7in]{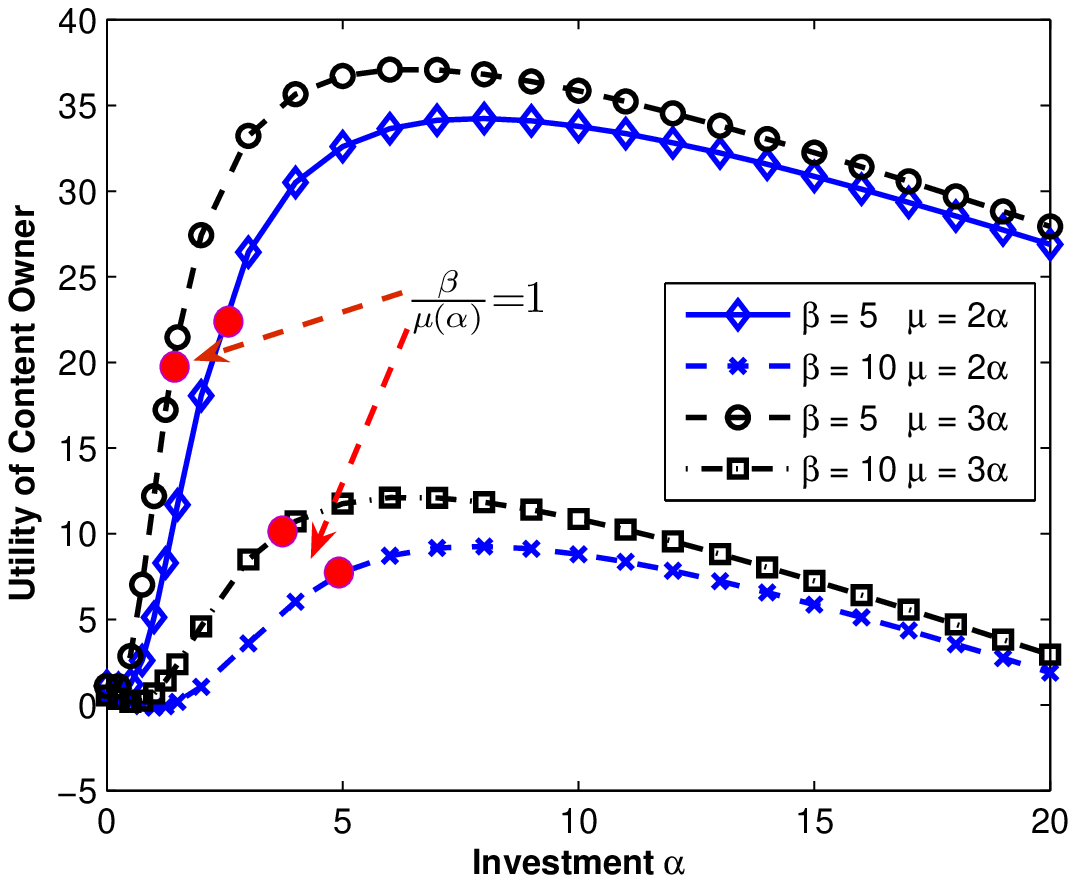}
   \caption{\small{Investment vs. utility of content owner.}}
   \label{fig:control1}
 \end{minipage}
\vspace{-0.5cm}
\end{figure*}

We now look at  the fraction of peers that will not acquire the file. We assume that $Y(0)=10$
and we recall that $N=300$.
Fig. \ref{fig:lambda_nofile}  (resp. Fig.\ref{fig:mu_peernofile}) displays this fraction as a function of $\lambda$ (resp. $\mu$)
for two different values of $r$ ($r=1$ corresponding to $X_c(0)=290$ and $r=0.5$ corresponding to $X_c(0)=140$ and $X_f(0)=150$).
In each figure, both simulation  and mean-field approximation results are reported. The fraction of peers without the file is a decreasing
function of the pairwise contact rate $\lambda$ and an increasing function of the cooperation degre $\mu$.
The mean-field
approximation is obtained as the unique solution $x_c(\infty)$ in $(0,x_c(0)$  of equation (\ref{eq:sec4.1.3})
where the initial condition  of the ODEs (\ref{ode}) is given by  $(y_0, x_c(0), x_f(0))=(Y(0)/N, X_c(0)/N,X_f(0)/N)$.
In both figures we observe a remarkable agreement between the simulation and the mean-field results (relative errors
never exceed $2\%$ when all peers are cooperative ($r=1$) and never exceed $7\%$ when half of the peers are free riders ($r=0.5$)).
We also note that the fraction of peers without the file  considered as a function of $\lambda$ (resp. $\mu$) is larger (resp. smaller)
when $r=0.5$ than when $r=1$; this is  of course not surprising since,  unlike cooperative peers, free riders do not contribute to the file dissemination.

\subsection{Action against unauthorized file downloading}
We now evaluate the impact  of actions against unauthorized file downloading. For that, we use
the framework developed in Section \ref{sec:control}.   Since the simulations in \cite{TR10}  show that, for large
$N$,  $\overline{T}(\alpha)$ in Section \ref{sec:control} is a good approximation of the expected time, $T_N$, needed for
an arbitrary peer to get the file we only consider the utility function $h(\alpha)$ (see  (\ref{utility}) ). We assume  that the cooperation degree $\mu(\alpha)$
is given by $\mu(\alpha)=\mu\alpha$. There are 500 peers ($N=500$) at time $t=0$   including two persistent publishers ($Y^*=2$).
We assume that $Y(0)=0$ so that $X(0)=N-Y^*=498$.  The initial condition of the ODE (\ref{ode3}) is  $(y_0,x_0)=(0,X_N(0)/N)$
with $y^*=2/N$.   Fig. \ref{fig:control1} displays the mapping $\alpha \to h(\alpha)$, for two different values of $\beta$ and $\mu$.
We observe that  a small investment cannot obviously postpone the expected delivery delay of the file, resulting in a decreased
utility. As the investment grows, the utility of the content owner increases significantly.
The curves in Fig. \ref{fig:control1} also show how large an investment has to be to counteract P2P illegal downloading. Note the content owner
can still have an increased utility when the ratio $\beta/\mu(\alpha)$ is greater than one, as  the utility is maximized across all curves when the ratio
$\beta/\mu(\alpha)$ lies between two and three.

\section{Related work}
\label{sec:related}

\begin{comment}
The current Internet has witnessed the rapid development
of various P2P services such as BitTorrent for file downloading, PPlive and PPstream for realtime streaming,
peer-assisted software update etc. We are particularly interested in the P2P file sharing systems and give
a brief literature review.
\end{comment}

There has been a number of work on the mathematical studies of structured and unstructured
P2P-based content distribution. A seminal
work can be found in \cite{Infocom04:Yang}. The authors
propose a continuous-time branching process to analyze service capacity (i.e. maximum
rate of downloading) and a coarse-grain Markov model to characterize the steady state of downloading rate.
In \cite{Sigcomm04:Qiu}, Qiu and Srikant propose a fluid model composed of ordinary differential equations to describe
the dynamics of BitTorrent systems. Authors in \cite{IWQoS06:Fan} further propose a novel fluid model based on
stochastic differential equations. This new model also extends \cite{Sigcomm04:Qiu} to multi-classes system and
is able to describe chunk availability. Mundinger et al.\cite{Scheduling:Mundinger} propose a deterministic
scheduling algorithm to achieve the optimal makespan for a structured system which requires global
knowledge. A coupon model is put forward in \cite{Sigmetrics05:Massoulie} to investigate the effectiveness
of a generic P2P file sharing system. %Authors in \cite{Performance07:lin} present an improved model with tighter
%bounds subsequently.

Recently, the process of file dissemination has attracted a lot of attentions. Cl\'evenot et al. adopt a  hybrid
approach (fluid and stochastic) to analyze Squirrel, a P2P cooperative web cache in \cite{Infocom04:Philippe}.
In \cite{Si} Queija et al. study
the scaling law of mean broadcasting time in a closed P2P swarm with constant request rate.
Authors in \cite{Sigmetrics08:Simatos} formulate a ball-and-urn model to characterize the ``flash crowd'' effect in a
closed P2P networks.
The content provided by P2P networks such as music, movies and software are usually unauthorized. Content
provided are therefore inclined  to combat  illegal downloading/uploading via technical solutions. Authors of
\cite{sulan1} and \cite{sulan2} propose a M/G/$\infty$ queueing model to access the efficiency of
non-cooperative measures against unauthorized downloading. Authors in \cite{CoNext09:Sadoc} propose
a similar but elegant queuing model to study the impact of bundling strategy of file availability and downloading rate.
%Authors in \cite{Sigmetrics06:Kumar} propose fluid
%models to study the propagation of corrupted/polluted content as well as the impact of counter measures.
Our general model is inspired by the one in \cite{Si}. However, it
differs from \cite{Si,Sigcomm04:Qiu,Infocom04:Yang,sulan2} in four ways: 1) we are studying the transient
behavior; 2) a peer can initiate a number of random contacts, instead of one, with other peers;
3) we observe several phase transitions in response to system parameters; 4) we adopt Markov branching process
and mean-field approaches to characterize the file dissemination model comprehensively.

%Different from the work mentioned above, we focus on the dissemination process
%of unauthorized P2P files and the impact of counteractions.

\section{Conclusion}
\label{sec:conclude}
In this paper we have proposed to use the theory of continuous time
branching process as well as of the dynamics of epidemics in order
to study the transient behavior of torrents that occur in P2P systems.
The use of these tools allowed us to compute the probability of early
extinction of the torrent as well as the expected time until that extinction,
the availability of a file in the system, the maximum availability and
when it occurs, and the size of the torrent.
% (the fraction of nodes
%that are eventually infected).
This is used for analyzing the impact of
measures to decrease non-authorized Internet access to copyrighted files.
We identify regimes in which the performance measures are quite sensitive
to such measures and others in which the measures have very limited impact.
In particular, we present two counteractions against unauthorized file sharing
in the presence of illegal publishers.
Our methodology can be extended to analyze file bundling that serves as a positive action
of file dissemination.

\begin{comment}
Besides, our methodology
can be extended to various P2P applications, such as the content distribution
networks with publishers and the P2P file bundling.
\end{comment}

\end{document}